\documentclass[prl,reprint, superscriptaddress]{revtex4-1}
\usepackage[utf8]{inputenc}

\usepackage{natbib}

\usepackage{amsmath}
\usepackage{dsfont} 
\usepackage{esint}     
\usepackage{bm}
\usepackage{mathtools} 

\usepackage{qcircuit}
\usepackage{bbold}
\usepackage{hhline}
\usepackage{physics}

\usepackage{subcaption}
\usepackage{graphicx}
\usepackage[justification=Justified, format=plain]{caption}
\usepackage{svg}

\usepackage{appendix}
\usepackage[hidelinks]{hyperref}
\usepackage{xcolor}

\usepackage{tikz}  
\usetikzlibrary{positioning}
\usetikzlibrary{decorations.pathreplacing,calc}
\usetikzlibrary{arrows,shapes,snakes,
		       automata,backgrounds,
		       petri,topaths}				
\usetikzlibrary{positioning}
\usetikzlibrary{decorations.pathreplacing,calc}


    
\usepackage{ifthen} 

\usepackage{verbatim}

\usepackage{algorithm}
 \usepackage{algpseudocode}

\newcommand{\cov}{\bm{\sigma}}

\begin{document}

\title{\textbf{Gaussian-boson-sampling-enhanced dense subgraph finding shows \\ limited advantage over efficient classical algorithms}}
\author{Naomi R. Solomons}
\email{naomi.solomons@bristol.ac.uk}
\affiliation{Quantum Engineering Centre for Doctoral Training, Centre for
Nanoscience and Quantum Information, University of Bristol, Bristol BS8 1FD, United Kingdom}
\affiliation{Quantum Engineering Technology Labs, H. H. Wills Physics Laboratory and Department of Electrical and
Electronic Engineering, University of Bristol, Bristol BS8 1UB, United Kingdom}
\author{Oliver F. Thomas}
\affiliation{Duality Quantum Photonics, 6 Lower Park Row, Bristol, United Kingdom, BS1 5BJ}
\author{Dara P. S. McCutcheon}
\affiliation{Quantum Engineering Technology Labs, H. H. Wills Physics Laboratory and Department of Electrical and
Electronic Engineering, University of Bristol, Bristol BS8 1UB, United Kingdom}
\affiliation{Duality Quantum Photonics, 6 Lower Park Row, Bristol, United Kingdom, BS1 5BJ}

\date{\today}

\begin{abstract}
    Recent claims of achieving exponential quantum advantage have attracted attention to Gaussian boson sampling (GBS), a potential application of which is dense subgraph finding. We investigate the effects of sources of error including loss and spectral impurity on GBS applied to dense subgraph finding algorithms. We find that the effectiveness of these algorithms is remarkably robust to errors, to such an extent that there exist efficient classical algorithms that can simulate the underlying GBS. These results imply that the speedup of GBS-based algorithms for the dense subgraph problem over classical approaches is at most polynomial, though this could be achieved on a quantum device with dramatically less stringent requirements on loss and photon purity than general GBS. 
\end{abstract}
\maketitle

\section{Introduction}

Gaussian boson sampling (GBS) is a non-universal model of quantum computation which samples from the photon number distribution of Gaussian squeezed states passed through a passive linear interferometer~\cite{hamilton2017gaussian}. Unlike universal quantum computers, GBS can be realised by currently available quantum devices at a scale which is not efficiently simulable by a classical computer, and hence has been the subject of early claims of quantum advantage~\cite{lund2017quantum, zhong2020quantum, madsen2022quantum}, challenging the extended Church-Turing thesis~\cite{shchesnovich2014conditions}. Recent proposals have suggested that GBS can be used for a number of different applications~\cite{bromley2020applications}, several of which involve examining properties of graphs. These include dense subgraph identification, a problem which occurs in computing~\cite{kumar1999trawling}, computational biology \cite{hu2005mining, habibi2010protein}, and finance~\cite{arora2010computational}, as well as being used to predict molecular docking configurations~\cite{banchi2020molecular}. 
Given the conjectured exponential complexity of simulating GBS, there is the potential that GBS enables genuine useful quantum computing applications offering exponential speedup.

However, current and near-term GBS experiments are likely to be significantly affected by error, having an impact on the effectiveness of GBS-based algorithms and the corresponding extent of any speedup over classical approaches. For example, in the presence of sufficient photon distinguishability and loss, GBS experiments can be efficiently classically simulated~\cite{renema2020simulability, qi2020regimes}. On the other hand, accurately modelling imperfect photon purity requires the simulation of multiple spectral modes 
increasing the simulation complexity, and the impact of spectral impurity is subsequently not well understood in the context of large scale GBS.  


This work uses methods for simulating GBS as described in \cite{thomas2021general}, applied to the stochastic densest $k$-subgraph (DkS) finding algorithms in \cite{bromley2020applications, arrazola2018using}. We examine how effective GBS experiments are for this application, when including the effects of spectrally impure sources and loss. The densest $k$-subgraph problem is known to be NP-hard in general, and as such an efficient quantum algorithm is unlikely (as it is widely thought that NP is not in BQP, the class of problems solvable by a quantum computer, so NP-hard problems cannot be efficiently solved by any quantum computer, let alone GBS). Nevertheless it is important to know how errors will affect quantum approaches, and to elucidate the origin of any speedups offered, polynomial or otherwise. We show that the DkS finding algorithms are extremely robust to these sources of errors, even in regimes that may be efficient to simulate classically. 
These results suggest that GBS applied to the DkS problem is likely to offer only a polynomial speedup over classical computing approaches at best, but that this advantage does not impose challenging hardware requirements on loss and photon purity, and could be realised more readily than a general GBS device. Given the level or errors that can be tolerated for these algorithms, we speculate whether any advantage offered on quantum devices should really be considered `quantum', or if `analogue' or `optical' may be more fitting terminology.

\section{Background}

\subsection{Gaussian boson sampling}

A Gaussian state is defined as a quantum state in which the Wigner function $\mathcal{W}(\textbf{p}, \textbf{q})$ is Gaussian, and thus an $m$-mode Gaussian state can be completely defined by a $2m \times 2m$ covariance matrix $\cov$, and a $2m$ displacement vector $\vec{D}$ \cite{adesso2014continuous}. In the following, we will assume $\vec{D} = \vec{0}$. 

Gaussian boson sampling consists of measurements in the Fock basis of an input Gaussian squeezed state  passed through an interferometer. The measurement probabilities for photon number resolving (PNR) detectors are given by~\cite{hamilton2017gaussian}:
\begin{equation} \label{eq: probability}
    P(S;\{s_i\}) = \frac{2^N} {s_1! s_2! ... s_n! \sqrt{|\sigma_Q|}} \text{Haf}(\mathcal{A}_S),
\end{equation}
in which $S$ is the subset of modes involved in the detection outcome, $s_i$ is the number of photons measured in mode $i$, $N = \sum_i s_i$, and $\sigma_Q = \sigma + \mathds{1}$. Given the matrix $\mathcal{A} = (X \otimes \mathds{1})(\mathds{1} - 2\sigma_Q^{-1})$, we construct $\mathcal{A}_S$ by repeating the $i$'th row and column of $\mathcal{A}$ according to $s_i$. 

The difficulty of classically simulating GBS is a result of the difficulty of calculating the matrix hafnian: 
$\text{Haf}(A) = \sum_{\mu \in M} \left( \prod_{k = 1}^{|S|} A_{\mu_{2k-1}, \mu_{2k}} \right)$, where $M$ is the set of perfect matchings of $S$, the different ways of `pairing' the indices (every permutation in which $\mu_{2k} < \mu_{2(k+1)}$ and $\mu_{2k} < \mu_{2k + 1}$). Each measurement carried out in a GBS experiment draws a sample from the distribution described by Eq. \ref{eq: probability}. Generating samples according to this distribution by direct calculation is $\#$P-hard \cite{deshpande2022quantum}. In the ideal case, the best algorithm for simulating GBS scales with the number of modes $m$ and the measured number of photons $N$ as $O(m N^3 2^{N/2})$ \cite{quesada2022quadratic}. The complexity of drawing samples from an experimental apparatus is linear in the number of samples, but alongside the difficulties of physically constructing the device, there is the initial computational overhead of calculating the correct interferometer settings to produce the desired output. Matrix decompositions, e.g. Williamson and Bloch-Messiah, are needed to find the transformations to generate the correct state; these generally require $O(m^3)$ time in practice~\cite{vasudevan2017hierarchical}.

\subsection{Mapping graphs to Gaussian states}

A graph $G$ is defined by a collection of vertices $V$ and connecting edges $E$, and is characterised by an adjacency matrix $A$. Here we restrict ourselves to equal-weight, undirected graphs, although all graphs can be represented as Gaussian states~\cite{walschaers2018tailoring}. To find this state we use the procedure defined in Ref.~\cite{bromley2020applications}. Firstly, the adjacency matrix $A$ is diagonalised find the eigenvalues $\{\lambda\}$. Next we pick the scaling parameter 
$c$ such that $c < \lambda_{\max}^{-1}$. We then construct $\mathcal{A} = c (A \oplus A)$, and the covariance matrix can be found using 
\begin{equation}
    \cov = 2 (\mathds{1} - X \mathcal{A})^{-1} - \mathds{1,}
    \qquad \mathrm{with} \qquad
    X=\begin{pmatrix}0&\mathds{1}\\\mathds{1}&0\end{pmatrix}.
    \label{eq:covfromgraph}
\end{equation}
This means that the probability of sampled outcomes will be proportional to $|\text{Haf}(A)|^2$ (as $\text{Haf}(A\oplus A) = |\text{Haf}(A)|^2$). The scaling parameter ($c$) ensures $\cov$ corresponds to physical squeezing values, and can be chosen to optimise the photon number distribution \cite{banchi2020molecular}. In principle the expected photon number can be arbitrarily high, but in practice this is limited by the amount of squeezing possible. The largest eigenvalue of the adjacency matrix is bounded above by the largest vertex degree in the graph, hence more well-connected graphs are likely to require greater squeezing levels \cite{minc1974nonnegative}.

The algorithms described in the next section are intended for use in the collision-free regime, which means that any outcomes containing multiple photons in one mode have negligible probability and can be ignored. For this reason we consider the use of threshold detectors which do not distinguish between photon numbers greater than zero. However, higher values of $c$, which result in more collisions, tend to favour more samples being drawn in the preferred photon number or click subspace for D$k$S, and hence improve the efficiency of the algorithm. As the size of the graphs increases, the probability of collisions decreases, hence allowing data collected in this work to use higher values of $c$.

\subsection{Dense subgraph identification}

We will consider densest $k$-subgraph identification, the problem of finding the subgraph of $k$ vertices with the largest density within the input graph $G$. The density of $G$ is given by:
    $\rho(G) = \frac{2 \abs{E(G)}}{\abs{V(G)} (\abs{V(G)} - 1)},$
in which $\abs{V(G)}$ and $\abs{E(G)}$ represent the number of vertices and edges in the graph, respectively. The maximum possible density of a graph is therefore $1$, for a fully connected graph (a clique). The D$k$S problem is NP-hard \cite{bhaskara2010detecting}, although solutions to variations on this problem can be found in polynomial time~\cite{gallo1989fast, tsourakakis2015k}, as can approximate solutions~\cite{feige2001dense}.

Consider a GBS experiment, where the Gaussian state leaving the interferometer has a covariance matrix as defined in Eq.~\ref{eq:covfromgraph} for some graph $G$. We can then associate subsets $S$ of output modes, corresponding to measurement outcomes, to subgraphs of $G$, where the modes in $S$ correspond to the vertices in the subgraph. As shown in \cite{arrazola2018using}, the number of perfect matchings in the graph, and hence the size of the hafnian of the adjacency matrix, is strongly correlated with the number of edges in a graph, and hence more dense subgraphs are the most likely sampling results. GBS can therefore used to seed algorithms that identify the densest subgraphs.

\section{Methods}

\subsection{Boson sampling generated subgraphs}

Different classical and quantum-enhanced algorithms exist for the DkS problem. In this work, we focus on the simplest approach, where the classical case involves sampling from the uniform distribution of $k$-vertex subgraphs, and the quantum-enhanced case takes samples using GBS and retains only those subgraphs of the correct size (the random search algorithm in \cite{arrazola2018using}), using these to select subgraphs. We tune the scaling parameter $c$ to maximise the probability of click patterns of the correct size. Our approach can therefore be considered a best-case scenario for the GBS algorithm, where no overhead in required number of samples is incurred to acquire $k$-vertex subgraphs. 
In the Supplementary Material we explore the performance of the base (photon number varying) GBS algorithm, and also compare to the performance of simulated annealing algorithms~\cite{arrazola2018using}. We use as a benchmark a deterministic classical algorithm, consisting of iteratively removing the lowest-degree vertex~\cite{asahiro2000greedily}, which does not always find the densest subgraph, but is guaranteed to find one of a density within a reasonable approximation ratio.

Our main text focuses on one representative, randomly generated graph, with further examples (showing qualitatively similar results) considered in the Supplementary Material. To create the graphs in this work, we use the Erd\H{o}s-R\'enyi form \cite{erdos1959publicationes} for generating a random graph $G(M,\rho)$, of size $M$ with density $\rho$, as described in the Supplementary Material. The quantum Gaussian optics emulator used to model outcome statistics is available online~\cite{qgot}.

\subsection{Modelling spectral impurities}

We introduce a new method for inserting internal degrees of freedom into a Gaussian state which represents a specific graph. Our method is general enough to work for any physical states, and can be thought of as replacing the idealised single (spectral) mode squeezers used to generate the state with more physical sources which have imperfections. The procedure to substitute in imperfect sources is as follows: 

\begin{enumerate}
\item Construct a graph $G$ with adjacency matrix $A$. 
\item Choose a scaling value, $c$, and construct a quantum state, $\cov$, from the graph, according to Eq. \ref{eq:covfromgraph}. 
\item Perform the Williamson decomposition to find the symplectic transformation that generates the state $\cov = \bm M \mathds{1} \bm M^\dagger$, $\bm M = \cov^{1/2}$ (for a pure state). 
\item Perform the Bloch-Messiah decomposition on the symplectic matrix to get the unitaries and single mode squeezers, $\bm M = \bm U \bm M^D \bm V^\dagger$. 
\item Replace the unitaries $\bm U$ and $\bm V$ with the full sized versions including spectral modes, $\mathcal{U} = \bm U \otimes \mathds{1}_{N_F}$, $\mathcal{V}^\dagger = \bm V^\dagger \otimes \mathds{1}_{N_F}$ (in which $N_F$ is the number of spectral modes).
\item Replace the set of single mode squeezers with a set of single spatial mode but multiple spectral mode squeezers $\bm M^D \mapsto \mathcal{M}^D$. These multiple spectral mode squeezers are normalised to the same amount of squeezing as the single spectral mode. Detectors are used that trace over the spectral modes, i.e. non-frequency resolving detectors. 
\item The multimode state can then be constructed using $\mathbb{\cov} = \mathcal{M} \mathcal{M}^\dagger$, where $\mathcal{M} = \mathcal{U} \mathcal{M}^D \mathcal{V}^\dagger$. To measure photons in the multimode formalism, we follow~\cite{thomas2021general}. 
\end{enumerate}

In order to characterise the sources, we first find the non-zero Schmidt coefficients. To do this, we  pick a set of basis functions, typically the Hermite polynomials for integrated optics, although any complete set of normalised functions is suitable. We then construct the vector of singular values $\{S\}$, using a set of coefficients, $X = \{x_i = s_i^2\}$. Remarkably, this allows a large number of geometric type distributions of the set of $X$, which can be parameterised by only two numbers -- here we use the labels $l$, $b$. We define the set $X$ as
$X_{l,b} = \{x_1\} \cup \{ x_i = k_{l,b}(i) (1-x_1), 2 \leq i \leq l\} \, ,$
so that by construction the normalisation condition on the values $k_{l,b}(i)$ is satisfied, $\sum_i k_{l,b}(i) = 1$. One useful form for the set $k_{l,b}$ is the geometric scaling,
$k_{l,b}(i) = b^{l-i}/(\sum_{j=1}^{l-1} b^{j-1}).$
Here $l$ is the number of non-zero Schmidt coefficients and $b$ is the base which sets the factor for the geometric scaling between the elements $x_i$. More information on this procedure is given in the Supplementary Information.

\section{Results}

\begin{figure}[ht!]
     \centering
        \includegraphics[width=0.48\textwidth]{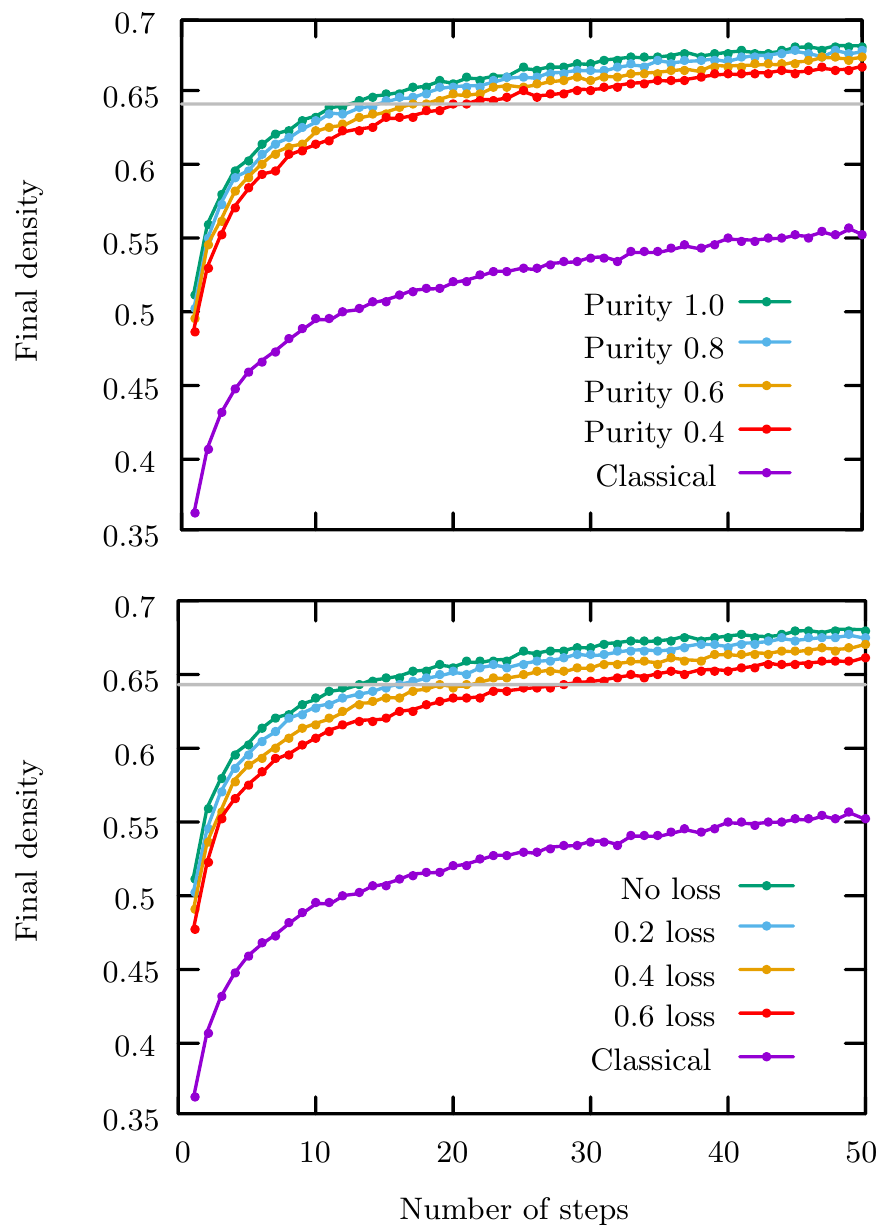}
     \caption{Performance of the classical (purple) and quantum-enhanced random sampling algorithm (green), in the presence of varying amounts of loss (top) and spectral impurity (bottom) as indicated. The grey line shows the performance of the deterministic classical algorithm~\cite{asahiro2000greedily}. The final maximum (density) of sampled subgraphs is shown as a function of sample batch size.}
     \label{fig: RS with error}
\end{figure}

The first graph we consider has $n=24$ vertices and a density of 0.362. The Supplementary Material also considers a graph with density 0.196, and a modified version containing a clique. We run the algorithm searching for the densest subgraph of $k=8$ vertices, which has density 0.714. 
Fig.~\ref{fig: RS with error} compares the performance of the classical (purple) and quantum-enhanced random sampling algorithms in the error free case (green). Shown also by the grey grid line is the result of the deterministic classical algorithm. 
The number of steps on the x-axis refers to the number of random subgraphs drawn from the target $k$-click subspace, with the final density being the maximum. 
The results are averaged over 1000 iterations.
We can see that the quantum-enhanced algorithms outperform the classical algorithms, similarly to ~\cite{arrazola2018using}. The deterministic algorithm performs well, but is quickly overtaken by the quantum-enhanced algorithms. 
Shown also is the GBS algorithm including different loss rates; Fig.~\ref{fig: RS with error} (top) as indicated. We do this by (after constructing the multi-spectral mode state) applying uniform loss to all modes. We also vary the spectral purity of the sources used in the GBS emulation using the method described above, shown in Fig.~\ref{fig: RS with error} (bottom). Interestingly, the quantum advantage offered by the GBS algorithm appears to be remarkably robust to loss and spectral impurity errors.

\begin{figure}
         \centering
         \includegraphics[width=0.45\textwidth]{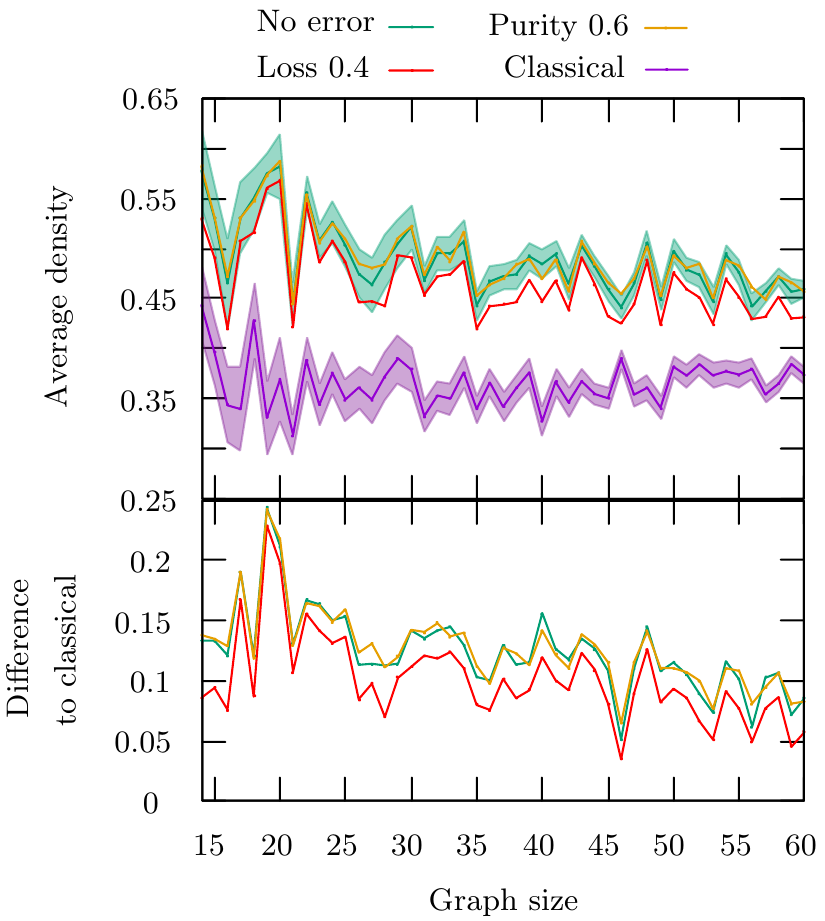}
        \caption{Top: Average density of sampled $\sqrt{n}$-subgraphs as a function of graph size $n$ with initial density $0.4$. The variance is shown (shaded region) for the classical and quantum case with no error. Bottom: difference between the average density of sampled subgraphs, in the quantum case, compared to the classical case.}
        \label{fig:scaling n}
\end{figure}

In order to see how the quantum advantage scales, we now investigate graphs of different sizes, all generated using the Erd\H{o}s-R\'enyi form \cite{erdos1959publicationes} with $\rho = 0.4$. 
We draw samples corresponding to $k$-vertex subgraphs where $k$ is the closest integer to $\sqrt{n}$, either from the uniform distribution (for the classical case) or from a simulated GBS device. Fig.~\ref{fig:scaling n} compares the average densities of these subgraphs after drawing 1000 samples, including with loss and spectral impurity. In the bottom plot, showing the difference between the quantum samples drawn and the classical samples, we see that the extent of the quantum advantage here appears to decrease with increasing graph size with a potential plateau approached above $50$.

\section{Discussion}

\begin{figure}
     \centering
     \begin{subfigure}[b]{0.45\textwidth}
         \centering
\includegraphics[width=\textwidth, trim={0 1.7cm 0 0},clip]{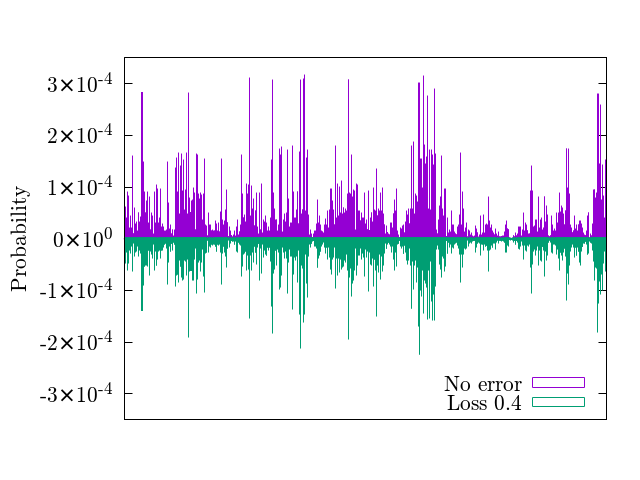}
     \end{subfigure}
     \begin{subfigure}[b]{0.45\textwidth}
         \centering
\includegraphics[width=\textwidth, trim={0 0 0 1.7cm},clip]{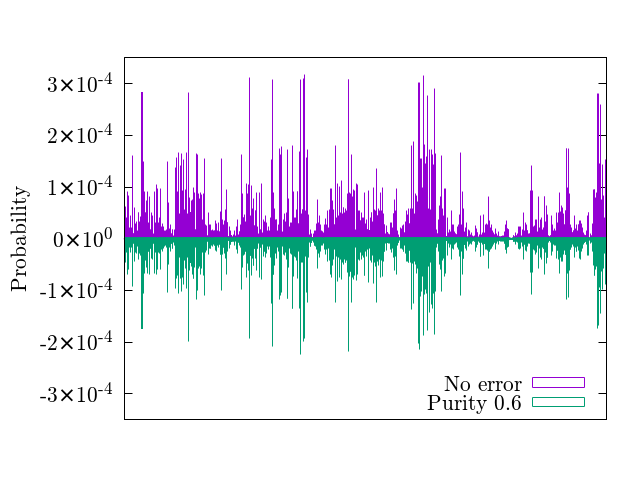}
     \end{subfigure}
     \caption{Probability distribution within the 8-click subspace, in the ideal case (purple) and with loss (green, top) and spectrally impure sources (green, bottom).}
     \label{fig: distributions}
\end{figure}

It is clear that the quantum-enhanced D$k$S algorithms are resilient against the forms of error considered. A considerable speedup (compared to the classical case) is shown by the examples with high levels of error (up to 60\% loss and sources with purity 0.4), with little difference between the effectiveness of these and the simulations with no error. This can in part be understood from Fig. \ref{fig: distributions}, which shows the outcome probability distributions within the 8-click space (the samples used in Fig.~\ref{fig: RS with error}) in the presence of loss and spectral impurity. Although some subtle differences can be seen in the distributions, the overall structure appears unchanged -- namely, that samples with the high probabilities retain high probabilities. This gives some intuition as to why the algorithms continue to select dense subgraphs in the presence of errors.

It is worth noting that we have considered the performance of the quantum-enhanced algorithm when samples have been selected from the appropriate photon number subspace. This reflects the fact that in realistic implementations, samples drawn outside of this subspace can be manipulated into graphs of the correct side, with minimal overheads (for example, by removing low degree vertices, adding vertices by combining graphs from different samples).  As such the postselected implementation presented here effectively represents the best-case scenario for the quantum algorithm. Without this postselection the speedup remains but is reduced as more samples need to be drawn. This is considered further in the Supplementary Material. 

When varying the graph size $n$, we see that the same pattern holds, in which simulations with loss and spectrally impure sources perform well. It also seems that the speedup of the quantum algorithm is smaller with larger $n$, and indicates a potential plateau at larger graph sizes. In addition, it can be seen that low spectral impurity has less of an impact than increased levels of loss. The variance of subgraph density in these samples is, in general, lower than that for the case with loss. This suggests that the experiment in this case may occasionally not be able to sample the densest subgraph, but also avoids sampling very low density subgraphs.

GBS with a high level of photon distinguishability is known to be efficient to simulate classically~\cite{renema2020simulability}. Similarly, increasing the spectral impurity approaches the limit of simulating thermal states, which are easier to simulate~\cite{rahimi2015can}, and hence reduces the usefulness of quantum resources. For a sufficient level of loss, GBS permits an efficient classical simulation~\cite{qi2020regimes}. Following the analysis of Ref.~\cite{qi2020regimes}, with loss of at least 43.7\%, the GBS simulated here is efficiently classically simulable. In the present case as loss is increased, we increase the squeezing to compensate in order to maximise the likelihood of $k$-click events (as described in the Supplementary Material). This increases the loss threshold above which the GBS is classically simulable, up to a maximum of 47.3\% for these parameters. As such, although the algorithms appear robust to errors, the level of errors means that efficient classical algorithms exist which perform the same underlying sampling task. We conclude that the apparent quantum advantage of GBS applied to the DkS is at best polynomial in run time.


The levels of loss and impurity tolerable suggest that these algorithms are not fully exploiting truly quantum mechanical effects such as quantum interference and entanglement. We note that it is not straightforward to engineer a graph problem requiring such phenomena since not all Gaussian states necessarily correspond to graphs in the manner described, i.e. adjacency matrices with real entries. Our results suggest future work could examine whether our findings are due to the structure of Gaussian states described by graphs, and if, for example, this is due to the limited treewidth of randomly produced graphs~\cite{oh2022classical}. Our investigations have found no examples of graphs corresponding to Gaussian states that are not robust to loss and impurity errors. The code used here is open source and can be run with a variety of different graph types~\cite{qgot}. 



\section{Conclusion}

We have used new methods of simulating Gaussian boson sampling to show that quantum-enhanced dense subgraph finding is particularly robust to spectral error and loss. 
This suggests that efficient classical methods exist that can implement the same algorithm as the quantum device, albeit with a potential polynomial overhead from simulating drawing samples. The implication is that a quantum device with modest requirements on loss and spectral purity could be used for the dense subgraph problem with little loss in performance, though this in itself raises questions as to whether the advantage gained can be said to be `quantum' in the sense that it is generally understood when applied to algorithms. 
This work has focused on studying Erd\H{o}s-R\'enyi random graphs, 
and further work could be done to generalise these results to different graphs with potentially more elaborate structures. Furthermore, more study is needed into the Gaussian states generated from graphs, and whether the conclusions drawn here apply to other applications of GBS. 
These results favor ongoing speculation as to the possibility of efficiently simulating GBS within certain regimes, in particular whether it is possible to efficiently calculate hafnians of matrices with all positive values.

\section{Acknowledgements}

We would like to thank Patrick Yard and Anthony Laing for feedback on the manuscript, and Ryan Mann for useful discussions. N.R.S. is supported by the Quantum Engineering Centre for Doctoral Training, EPSRC grant EP/SO23607/1.

\bibliography{references}
\bibliographystyle{apsrev4-1}

\onecolumngrid

\begin{appendices}

\section{Calculating the scaling parameter}

Here we discuss how to calculate the optimal scaling parameter, $c$, in order to maximise the probability of drawing samples in the $k$-click subspace. We first eigendecompose the adjacency matrix $A = U \Lambda U^T$, with $\Lambda = \text{diag}\{|\lambda_i|\}$, in which $\lambda_i$ are the eigenvalues of $A$. We then rescale to give the diagonal matrix $t_D = c\Lambda$ with $t_D=\mathrm{diag}\{\tanh r_i\}$ with $r_i$ the squeezing in mode $i$. When using threshold detectors, the expected number of `clicks' (detectors registering at least one photon during a measurement) is given by
\begin{equation}
    \langle \hat{C} \rangle = M - \sum_{i=1}^M P_{\text{vac}}(i), 
\end{equation}
where $M$ is the number of modes, and the vacuum probability in mode $i$, $P_{\text{vac}}(i)$ is given by:
\begin{equation}
    P_{\text{vac}}(i) = \left( \left| \sum_j |U_{ij}|^2 \left( \frac{1 - l t_D^2}{1 - t_D^2} \right)_j \right|^2 - (1-l)^2 \left| \sum_j U_{ij}^2 \left( \frac{t_D}{1 - t_D^2} \right)_j \right|^2 \right)^{-1/2}.
\end{equation}\label{eq: calculating c}
Increasing $c$ means increasing the squeezing parameter, and hence this increases the likelihood of collision events, which makes the GBS experiment more simulable \cite{martnez2022classical}. The effect of squeezing on dense subgraph finding was investigated experimentally in \cite{sempere2022experimentally}.

\section{Generating random subgraphs}

 Our implementation of the Erd\H{o}s-Rényi model \cite{erdHos1960evolution} is as follows:
\begin{enumerate}
    \item Create a set of vertices $V$ of size $M$.
    \item Start at the first vertex $v_{i=1}$ in $V$.
    \item For every other vertex $\{v_j \in V \setminus v_i \}$, pick $r_j \in [0,1]$ from a uniform distribution.
    \item If $r_j < \rho/2$ then add an edge between $v_i$ and $v_j$.
    \item At the last vertex, $v_{i=M}$, stop, else move to the next vertex, $v_i \leftarrow v_{i+1}$, and go to 3. 
\end{enumerate}

\section{Constructing arbitrary purity sources}

We construct the vector of Schmidt coefficients $\{S\}$, such that they obey the following normalisation and purity conditions (in which $P$ is purity):
\begin{align}
    &\sum_i S_i^2 = 1 \label{eq:snormalisation} \, , \,\,\, \sum_i S_i^4 = P := \Tr[\hat{\rho}^2] \, .
\end{align}
In order to simplify the two conditions in Eqs.~\ref{eq:snormalisation} we construct a new set of coefficients, $X = \{x_i = s_i^2\}$ which reduces the two conditions to:
\begin{align}
    &\sum_i x_i = 1 \, , \,\,\, \sum_i x_i^2 = P \, .
    \label{eq:xpurity}
\end{align}
We define the set $X$ (which are the Schmidt coefficients squared) as,
\begin{equation}
    X_{l,b} = \{x_1\} \cup \{ x_i = k_{l,b}(i) (1-x_1) | 2 \leq i \leq l\} \, ,
\end{equation}
so that, by construction, the normalisation condition is satisfied, where the condition on the values $k_{l,b}(i)$ ensures that they are normalised to 1, $\sum_i k_{l,b}(i) = 1$. One useful form for the set $k_{l,b}$ is the geometric scaling,
\begin{equation}
    k_{l,b}(i) = \frac{b^{l-i}}{\sum_{j=1}^{l-1} b^{j-1}} \, .
\end{equation}
Here $l$ is the number of non-zero Schmidt coefficients and $b$ is the base which sets the factor for the geometric scaling between the elements $x_i$. $i$ is the index of the element of the set of $X_{l,b}$ for the coefficient value.

Using these definitions the purity condition~\autoref{eq:xpurity} is then,
\begin{equation}
\begin{split}
    \sum_i x_i^2 &= x_1^2 + (k_{l,b}(2))^2 (1-x_1)^2 + \dots \\ 
    &+ (k_{l,b}(l))^2 (1-x_1)^2 = P \, ,
\end{split}
\end{equation}
and grouping the terms by the order of $x_1$ we see it produces a quadratic equation in $x_1$
\begin{equation}
\begin{split}
    \left( 1 + \sum_{i=2}^l (k_{l,b}(i))^2\right) x_1^2 &- 2 \left( \sum_{i=2}^l (k_{l,b}(i))^2 \right) x_1  + \left( \sum_{i=2}^l (k_{l,b}(i))^2 \right) - P = 0 \, ,
\end{split}
\end{equation}
which is then straightforward to solve. Since all the elements of $S$ are positive all the elements of $X$ are positive, and as such the positive root can be taken. Lastly, we calculate the remaining $x_i$ by substituting $x_1$ back in: 
\begin{equation}
    x_i = k_{l,b}(i) (1 - x_1) \, ,
\end{equation}
and the set $S$ is the (positive) square root of $X$,
\begin{equation}
    S = \{\sqrt{x_i} | x_i \in X \} \, .
\end{equation}
Due to the choice of the two parameters $l$ and $b$, for a particular set of $S$ with the desired purity there are many distributions of Schmidt coefficients that generate a source with the same purity. Also we note that our construction does not generate all Schmidt distributions but it can be used to generate a large class of distributions.

One interesting consequence of this particular form for $X$ is that for any purity with $\frac{1}{2} \leq P \leq 1$ only two terms are needed. For quantum information processing applications we are typically interested in sources with purities close to $1$, and as such the majority of physically relevant or interesting sources can be described with two Schmidt modes. This offers a significant simplification when including a large number of spectral degrees of freedom into a model. Although including spectral modes that are traced over in the detection calculations only introduces a cubic overhead~\cite{thomas2021general}, these factors are still large. Using this relation we can, when required for speed, drastically reduce the number of spectral modes modelled from some large $l$ to only 2. 

\section{Further graph examples}

\begin{figure*}[htb]
     \centering
     \begin{subfigure}[h!]{0.4\textwidth}
         \centering
         \includegraphics[width=\textwidth]{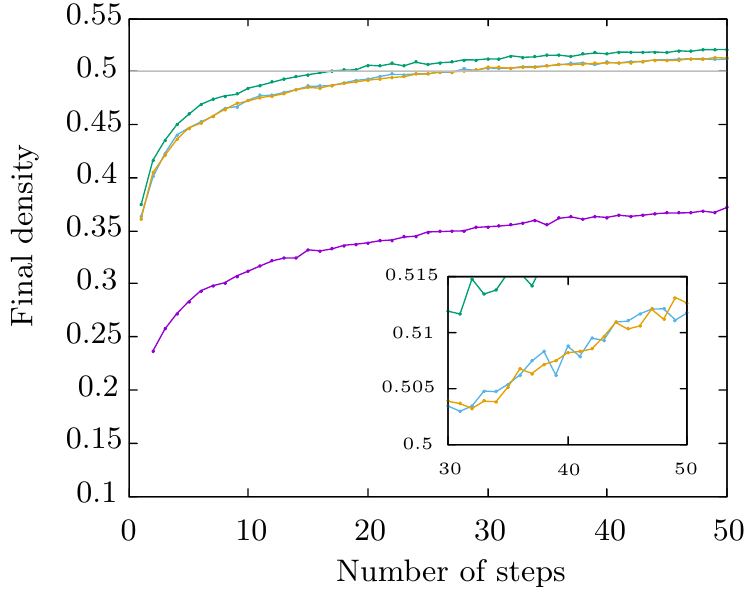}
         \caption{Initial graph has density 0.196.}
     \end{subfigure}
     \hfill
     \begin{subfigure}[h!]{0.55\textwidth}
        \centering
        \includegraphics[width=\textwidth]{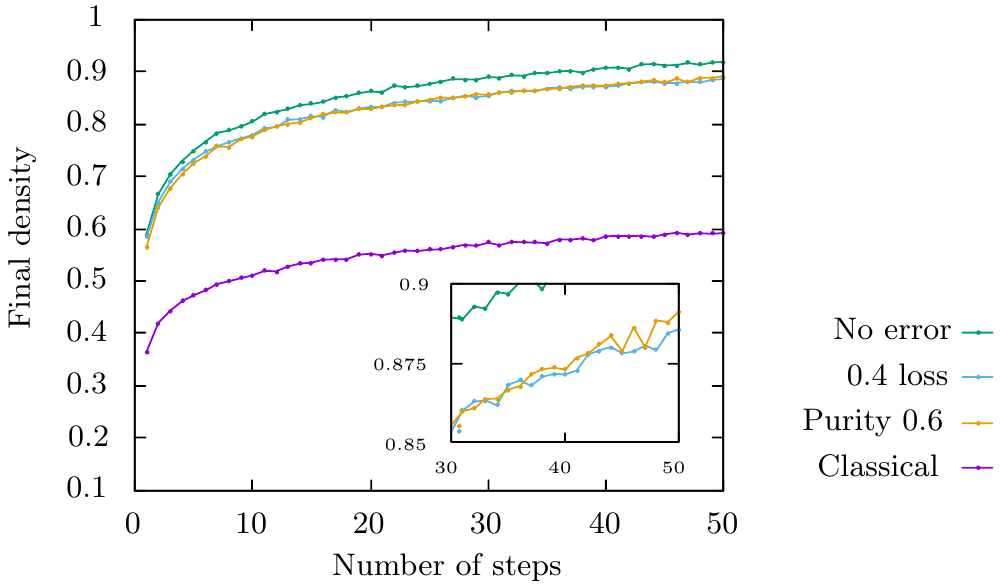}
        \caption{Initial graph has density 0.362 and contains a clique.}
        \label{fig:clique RS loss}
     \end{subfigure}
        \caption{Comparison of performance of the classical (purple) and quantum (green) enhanced random sampling algorithm, with different sources of error. The grey line shows the performance of the deterministic classical algorithm \cite{asahiro2000greedily} (it is able to identify the clique). The inset focuses on the region between 30 and 50 steps.}
        \label{fig:other graphs}
\end{figure*}

\begin{figure}[ht!]
        \centering
         \includegraphics[width=0.5\textwidth]{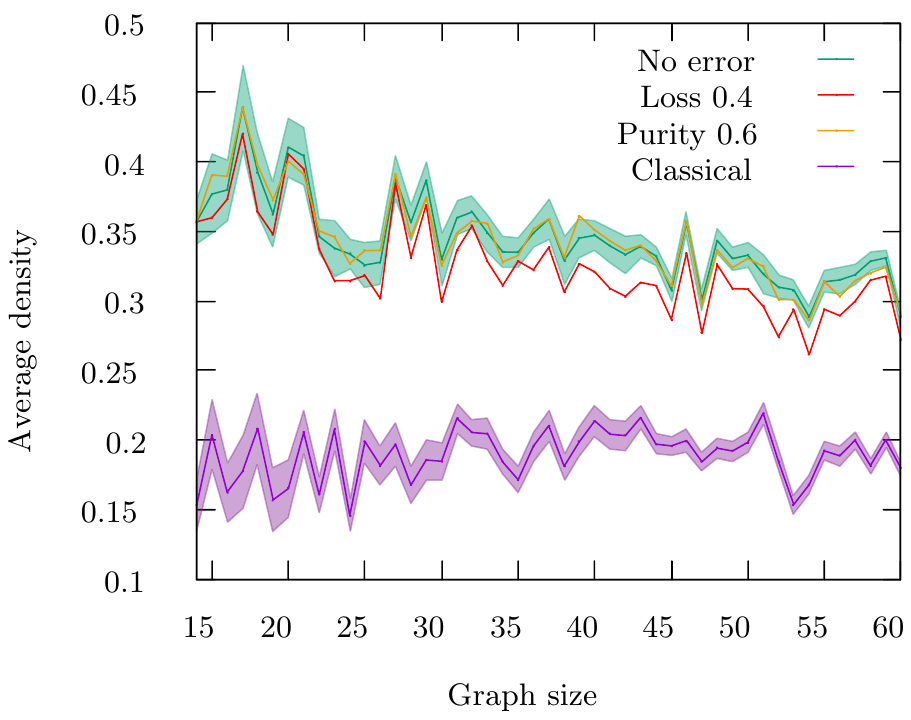}
        \caption{We plot the average density of subgraph found over 1000 samples, for different graph sizes, with initial density approximately 0.4. The variance is shown for the classical and quantum case with no error.}
        \label{fig:scaling n - lower density}
\end{figure}

To check the generalisability of our results we carry out the same analysis as in the main text but with with different  graphs. We show the performance of random sampling on two initial subgraphs, one of which is an Erd\H{o}s-Rényi graph with density 0.196, and one which is a modified version of the main graph from our paper, with the same initial density, but containing a clique (this is identified by the deterministic algorithm). Fig. \ref{fig:other graphs} shows the impact of loss and spectrally impure sources on dense subgraph finding applied to these graphs. Fig. \ref{fig:scaling n - lower density} shows the average density of graphs chosen by drawing samples from a GBS distribution, or by randomly selecting in the classical case, over 1000 repetitions. As in Fig. \ref{fig:scaling n}, this is done for a range of graph sizes, searching for subgraphs with size equal to the closest integer to $\sqrt{n}$.

These results suggest similar behaviour to the example discussed in the main text, being very robust to error and showing a significant speedup in the quantum case, even with high levels of loss and spectral impurity.

\section{Simulated annealing}

\begin{figure*}
         \centering
         \includegraphics[width=0.95\textwidth]{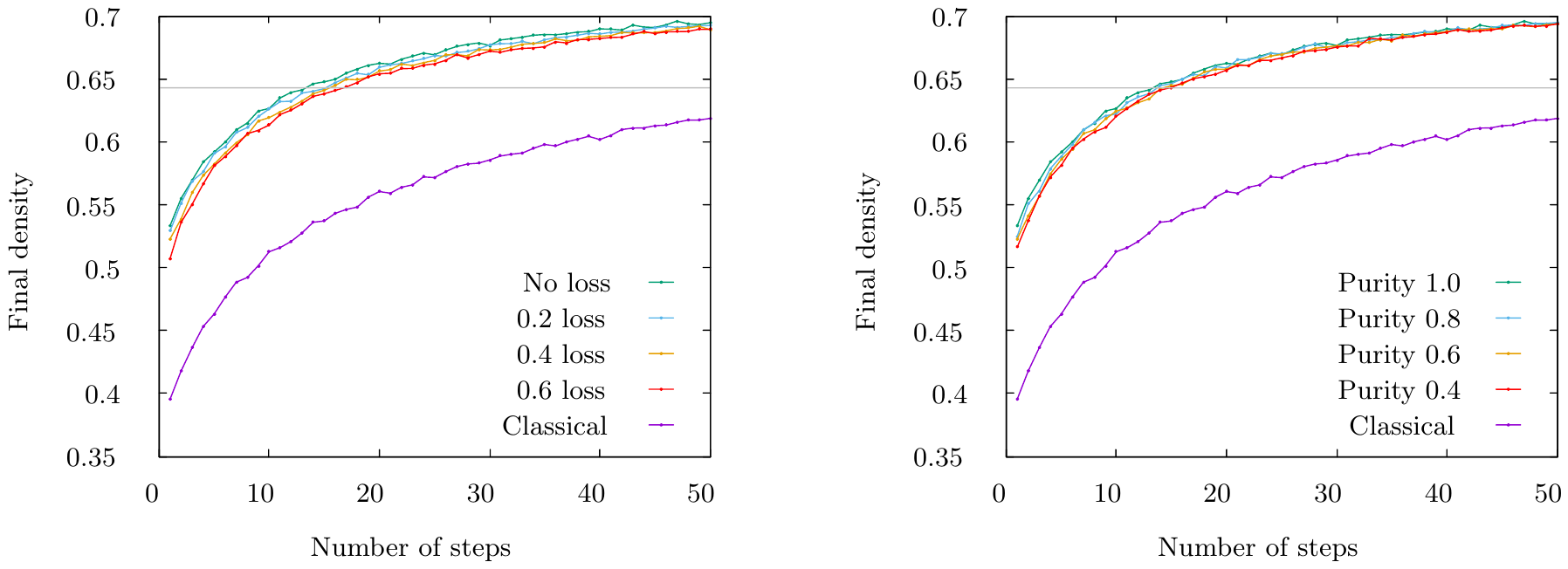}
        \caption{Comparison of performance of the classical and quantum enhanced simulated annealing algorithm, in the presence of varying amounts of loss (left) and with spectrally impure sources (right). The grey line shows the performance of the deterministic classical algorithm \cite{asahiro2000greedily}.}
        \label{fig: SA with error}
\end{figure*}

We now employ the quantum-enhanced simulated annealing algorithm, as described in \cite{arrazola2018using}, for finding dense subgraphs in our main example graphs (as introduced in the main text). Fig. \ref{fig: SA with error} shows the impact of loss (left) and impure sources (right). As in \cite{arrazola2018using}, this algorithm shows a smaller difference between the effectiveness of the classical and quantum-enhanced algorithms. Nonetheless, as with our previous results, this speedup remains even in the presence of significant levels of error.

Note that our implementation of the simulated annealing algorithm differs slightly from that described in \cite{arrazola2018using}. During the \textit{GBS-Tweak} subroutine, described in \cite{arrazola2018quantum}, a sample $T$ (in this case corresponding to a subgraph) is drawn, that facilitates an exploration of the search space, after drawing an initial sample $S$. We drew the sample $T$ from the modes that are not included in $S$, as opposed to the full set of modes. This was done to reduce run time while minimising impact on the performance of the algorithm. 
Here, `number of steps' (as indicated on the $x$-axis) refers to the number of steps in the \textit{GBS-Tweak} subroutine.

\section{Performance without postselection}

Previous results have considered the `number of steps' to only include detection events with $k$ click events. However, unlike the Aaronson-Arkhipov boson sampling scheme that uses single photons as input \cite{aaronson2013bosonsampling}, GBS does not have a fixed photon number at the detection stage. By choosing the scaling parameter $c$ appropriately, it is possible to bias the output towards a particular photon number (or number of clicks when using threshold detectors), but a spread of photon numbers will be observed in practice.

\begin{figure}
         \centering
    \includegraphics[width=0.7\textwidth]{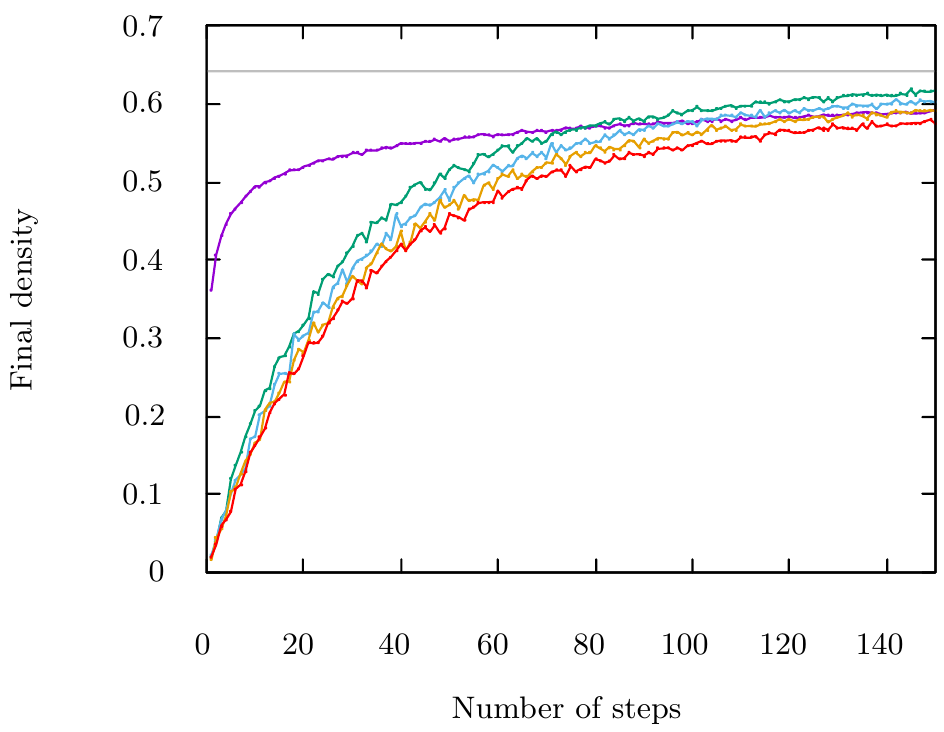}
        \caption{Performance of the classical and quantum-enhanced random sampling algorithm for dense subgraph finding, without postselecting GBS samples from the correct photon number subspace. The grey line shows the performance of the deterministic classical algorithm.}
        \label{fig:raw samples}
\end{figure}

In Fig.~\ref{fig:raw samples} we show the performance of the DkS random sampling algorithm, where here `number of steps' refers to the total number of samples drawn. We assume that any samples with the incorrect click number are discarded (realistically, these may be useful for local searches). We vary the loss parameter, and use the optimal scaling parameter $c$ (using the method previously described) for each different loss value (thus, we assume the error in the device is well understood and the squeezing can be increased to compensate). For the classical comparison, we repeat the previous method, and sample randomly across only the $k$-vertex subgraphs. As before, we can see that increased levels of loss do not have a significant impact on the performance of the quantum-enhanced algorithm.

\end{appendices}

\end{document}